# MIMO Based Multimedia Communication System

D. Kandar[1], Sourav Dhar[1], Rabindranath Bera[1] and C.K. Sarkar[2]

[1]Sikkim Manipal Institute of Technology, Sikkim Manipal University,Majitar, Sikkim, India 737132.
[2]Electronics and Telecommunication Engineering, Jadavpur University, Jadavpur, Kolkata, India 700032.

E-mail : r.bera@rediffmail.com, Phone: 09932778779; Fax: 03592-246112.

*Abstract:*
*High data rate is required for multimedia communication. But the communication at high data rate is always challenging. In this work we have successfully performed data chatting, Voice chatting and high quality video transmission between two distant units using MIMO adapter, Direct sequence spread spectrum system and MATLAB/SIMULINK platform.*

## I  INTRODUCTION

Communication technology requires to be sustainable in the sense of people's satisfaction for high data rate multimedia communication. Communication channel is the main constraint to achieve high speed data communication [1]. The third-generation (3G) and fourth-generation (4G) mobile communications are expected to provide a high-rate data services beyond 2 Mbps utilizing WCDMA, CDMA2000, OFDM and UWB [2]. The capacity improvement of wireless network has drawn considerable attention utilizing  Smart Antenna based multiple-input multiple-output (MIMO) technology. MIMO utilizes multiple antenna arrays at both the transmitter and receiver side of a communication link and significantly improves the capacity over the single antenna system [3].

Data rate for voice communication is reduced to the minimum level for additional users [5]. But people are interested about the performance of a system. For higher data rate, enhancement of transmit power will cost more because of the logarithmic relationship between the capacity of a wireless link and signal-to-interference-and-noise-ratio (SINR) at the receiver end [6].In a multi user environment the wireless transmission performance is fundamentally limited by interference due to the presence of signals from other users as well as from multipath. Specially, the performance of wireless communication systems is limited by time dispersion in the channel as a result of multipath. Dispersion gives rise to frequency-selective fading that results in a distorted frequency spectrum of the information signal[6]. It is usually assumed that the multipath delay spread is smaller than a symbol duration [7] because larger values of delay spread cause significant inter symbol interference (ISI), which limits the achievable data rate. Fading, Jamming, Security are also the important issues in wireless communication.

Matlab/Simulink is chosen as the platform for simulation. In MATLAB/SIMULINK, multimedia transmission is possible using the blocks of physical layer wireless communication like UDP HOST Send. 'UDP Send' sends a UDP packet to a remote interface identified by IP address & IP port parameters. 'UDP Receive' receives UDP packets from indicated IP address & IP port. A program is developed in Simulink to transmit & receive data , audio and  video .

## II  THE MIMO-HARDWARE IN LOOP (HALO)

DSSS system is explored first to achieve the higher data rate over secured mobile environment. The author and his group is successful in sending multimedia data through a IEEE 802.11b adapters using UDP mode of Matlab. A DSSS communication system made up of WiFi b PCI adapter fitted inside a PC acting as Transmitter and another DSSS communication system made up of another WiFi b PCI adapter fitted inside another PC acting as Receiver are operational at SMIT using 2.4 GHz radio carrier. The transmitter and Receiver are separated at a distance of 110 m [8]. The Block diagram of such communication system is shown in Fig 1.  The transmitter radiates carrier using a dish antenna of diameter 6 ft with a horn at its focus. The horn is connected to WiFi b adapter fitted inside a PC. At the other end the receiver with another WiFi b adapter and a second horn is placed.

The same experiment is repeated with the WiFi N card in HArdware-in-LOop (HALO) mode. A MIMO system is formed by replacing 'Wi-Fi b' card with 'WiFi N' card and putting 3 antennas at the radio units which results in a good interference as well as multipath rejection and higher SNR. Using wi-fi 'N' card (Shown in Fig. 2) hardware over the loop system has been implemented, where wireless loopback is performed between two pc's i.e. the transmitted video(from Tx End) has been Received and Transmitted (from Rx End) again from the distant PC and then Received and tested at the same PC(from Tx End). Here UDP protocol is used in the transport layer. Data chatting, Voice chatting and high quality video transmission between two distant units are performed as described above.

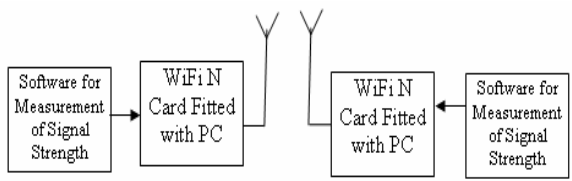

Fig. 1. Block Diagram of a DSSS Communication system

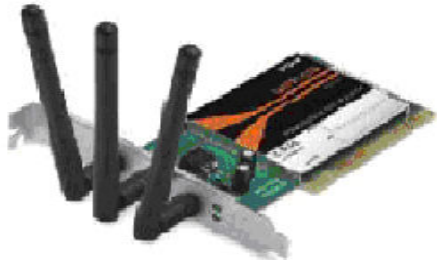

Fig. 2. PCI Slot Based WiFi N PCI adapter.

## III. SOFTWARE FOR HALO SYSTEM

Matlab/Simulink is used as the platform to implement this system. In MATLAB/SIMULINK, multimedia transmission is possible using the blocks of physical layer wireless communication like UDP HOST Send, UDP over wireless carrier by configuring TCP/IP of UDP port. 'UDP Send' sends a UDP packet to a remote interface identified by IP address & IP port parameters. 'UDP Receive' receives UDP packets from indicated IP address & IP port. We have developed a program in simulink to transmit multimedia data like audio, video & receive it.

*1. Transmitter Block*

The basic block diagram of the Transmitter (Tx) and Receiver (Rx) system is shown in Fig. 3 and Fig 4 respectively. The Tx block mainly consists of Audio, Video source and UDP block. The Audio and Video data are being sent in different ways using different UDP ports. Color conversion is required to send Video data. The Color Space Conversion block converts color information from R'G'B' to Y'Cb'Cr'. After color conversion the 'Chroma Resampling' block downsamples or upsamples chrominance components of pixels to reduce the bandwidth required for transmission of the signal. Byte pack block converts data from integer data type to a single uint8 vector. We have converted the video format to uint8 type as UDP block in the model supports only this data type. After byte packing the video data we have sent it through the wireless channel. Simultaneously Audio signal is also being processed. Unbuffer block unbuffers a M-by-N frame-based input into a 1-by-N sample-based output. Thus inputs are unbuffered row-wise so that each matrix row becomes an independent time sample in the output.

After unbuffering the audio source, we have sent the data using a different UDP port. Thus both video and audio could be transmitted simultaneously on the network. The multimedia data was passed from the application layer to the physical layer via MAC layer and the Data link layer. For the wireless clients, the data is passed to the wireless adapter and then to the antenna for transmission.

On the other hand, the data from the physical layer is passed to the UDP port and then transmitted on the network for the clients connected to the network through the Ethernet cable.

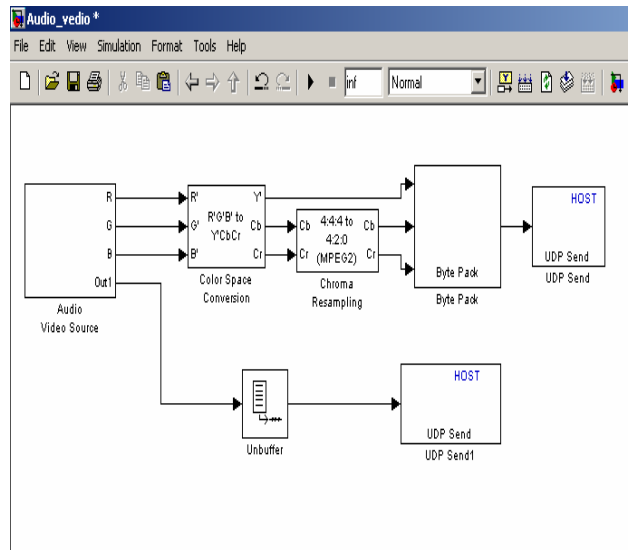

Fig. 3 The Simulink program for Multimedia data Tx System

*2. Receiver Block*

The destination is targeted and the transmitted data reaches the receiver through the network. The data is passed to a router and then to another router and so on till the receiver is detected. The data from the antenna of the receiver is sent to the physical layer of the UDP protocol that passes the data to the application layer through the datalink layer and the MAC layer. After this the model based to operate in the application layer does the reverse processing to get and decode the transmitted signal (Fig: 2). After receiving the video data using UDP port it has to be firstly unpacked and then converted to R'G'B' from Y'Cb'Cr'. Then we have connected a video viewer block to visualize the video data for its quality and the delay introduced in the system. This video viewer block enables to view a RGB video stream. Simultaneously we have buffered the received audio signal for audio visualization. In this way we have successfully sent the multimedia data over wireless channel efficiently between the clients. Complete transceiver block is shown in Fig. 5.

## IV. CONCLUSION

The DSSS system fails during multimedia video transmission since huge files are to be transmitted, but MIMO system is working satisfactorily under these circumstances. In MIMO communication system, the receiver enjoys the more or less constant average signal to noise ratio (SNR) of the received signal, whereas in conventional systems, which transmit all their energy over a single path, the received SNR varies considerably. The helical antennas were used for this experiment. Replacing helical antenna by directional antenna may increase the distance. The performance of the MIMO based multimedia system is so interesting that the authors dream to implement it in Intelligent Transportation System which is yet to be tested on moving car.

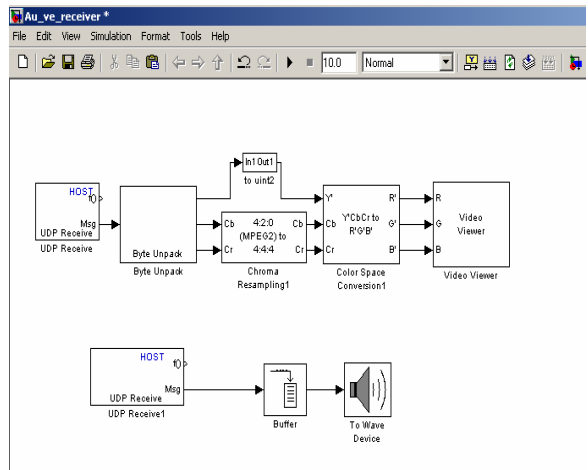

Fig. 4.The Simulink program for Multimedia data Rx System.

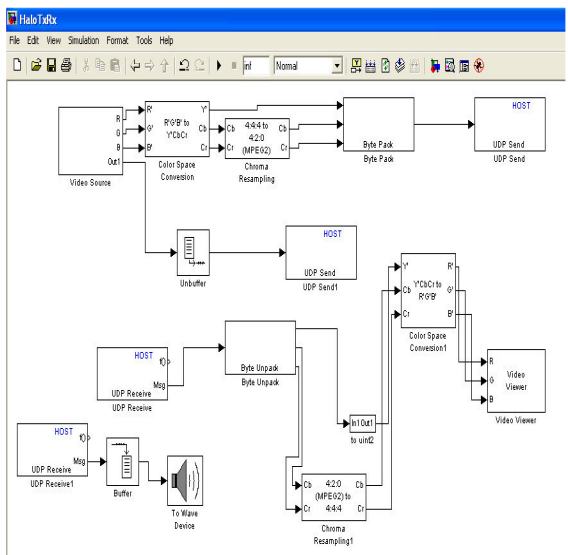

Fig. 5.The Simulink program for Multimedia data TxRx System

## V. ACKNOWLEDGMENT

Authors would like to thank all technical assistants of Electronics and Communication Department, SMIT for providing support both technically and financially.

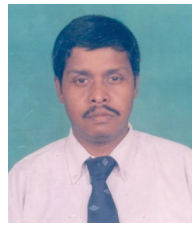

**Rabindra Nath Bera**: Born in 1958 at Kolaghat , West Bengal, INDIA. Received his B. Tech, M. Tech & Ph.D (Tech) from the Institute of Radiophysics & Electronics, The University of Calcutta, in the year 1982,1985 & 1997 respectively. Currently working as Dean (R&D) and Head of the Deparment, Electronics & Communication Engineering, Sikkim Manipal University,Sikkim, Microwave/ Millimeter wave based Broadband Wireless Mobile Communication and Remote Sensing are the area of specializations.

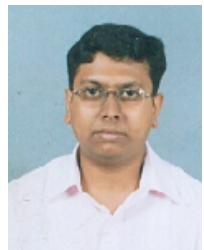

**Sourav Dhar:** Born in 1980 at Raiganj, West Bengal, INDIA. Received B.E from Bangalore institute of Technology and M.Tech from Sikkim Manipal Institute of Technology in the year 2002 and 2005 respectively. Currently working as a Senior Lecturer , Department of E&C Engineering , Sikkim Manipal University, Sikkim, India. Broadband Wireless Mobile Communication and Remote Sensing are the area of specializations.

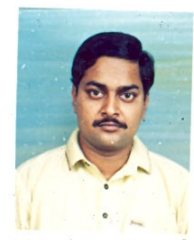

**Debdatta Kandar:** Born in 1977 at Deulia, Kolaghat, Purba Medinipur, West Bengal, INDIA. Received M.Sc. from Vidyasagar University and M.Tech from Allahabad Agricultural Institute-Deemed University in the year 2001 and 2006 respectively. Currently working as a Scientist-C, Department of E&C Engineering, Sikkim Manipal University, Sikkim, India. Broadband Wireless Mobile Communication and Remote Sensing are the area of specializations.